# Unsupervised Semantic Representation Learning of Scientific Literature Based on Graph Attention Mechanism and Maximum Mutual Information

Hongrui Gao[1], Yawen Li[2*], Meiyu Liang[1], Zeli Guan[1]

[1]Beijing Key Laboratory of Intelligent Communication Software and Multimedia, School of Computer Science (National Pilot Software Engineering School), Beijing University of Posts and Telecommunications, Beijing 100876
[2]School of Economics and Management, Beijing University of Posts and Telecommunications, Beijing 100876

**Abstract:** Since most scientific literature data are unlabeled, this makes unsupervised graph-based semantic representation learning crucial. Therefore, an unsupervised semantic representation learning method of scientific literature based on graph attention mechanism and maximum mutual information (GAMMI) is proposed. By introducing a graph attention mechanism, the weighted summation of nearby node features makes the weights of adjacent node features entirely depend on the node features. Depending on the features of the nearby nodes, different weights can be applied to each node in the graph. In addition, an unsupervised graph contrastive learning strategy is proposed to solve the problem of being unlabeled and scalable on large-scale graphs. By comparing the mutual information between the positive and negative local node representations on the latent space and the global graph representation, the graph neural network can capture both local and global information. Experimental results demonstrate competitive performance on various node classification benchmarks, achieving good results and sometimes even surpassing the performance of supervised learning.

## 1 Introduction

Currently, scientific literature resources are flooding the Internet [1][2]. How to extract important information from scientific literature and effectively represent them semantically is the core issue to realize the classification, retrieval, and recommendation of scientific literature.

Traditional methods mainly rely on expert experience to construct artificial features to represent scientific literature. For example, in academic information retrieval, literature titles and abstract texts are used to construct an inverted index [3]. But every time new text data is added, it is tedious to rebuild the entire index. In literature classification and recommendation systems, bag of words model [4], vector space model, and topic model [5] are used to build scientific literature vectors. However, the shortcomings of the bag of words model are obvious. When the vocabulary increases, there are at most ten words used for each sentence, which leads to the sparse matrix of each sentence, seriously affecting the memory and computing resources. Mainstream methods can map data into vector space and operate on vectors to complete specific tasks[6-9][13]. However, the existing vector space model-based document processing methods are based on word frequency information[10][11]. The similarity of two documents depends on the number of common words, and the semantic ambiguity of natural language cannot be distinguished.

Deep learning-based representation learning has received extensive attention recently[12][14-16]. The fundamental drawback is that the neural language model only focuses on the text semantics information of academic documents[17] while ignoring the relationship between academic documents. Some researchers solve the problem of imperfect data characteristics by means of multi-agent [18-19]. More and more researchers are fusing different features to better complete deep learning tasks[20-24].

In view of this, graph neural network [25] is proposed to extract the relationship structure information between documents from the document citation network and fuse them with the semantic information of the document text, so as to construct the representation vector of academic documents. However, most of the existing studies use supervised graph neural networks to learn feature representations of documents[26][27], which have following two drawbacks: For specific tasks, supervised graph neural networks need to develop a huge amount of excellently labeled data; The feature representation of text obtained by supervised graph neural network[28]is highly coupled with the task of labeling datasets, and it is difficult to directly transfer to other tasks, resulting in poor universality of feature representation.

Compared to supervised learning method, unsupervised graph neural networks perform better. Because they can directly learn general document feature representations from unlabeled document network data[29].

Based on this, this paper proposes an unsupervised semantic representation learning method for scientific literature based on graph attention mechanism and maximum mutual information (GAMMI). The following are this paper's main contributions:

1) A semantic representation learning method for scientific literature based on graph attention mechanism and maximum mutual information is proposed. By introducing graph attention mechanism, the relationships between nodes features are better included into the model, and the node representation is only related to adjacent nodes, which can be directly applied to inductive learning without receiving all graph information.

*Corresponding author: Yawen Li (warmly0716@126.com).

2) An unsupervised graph contrastive learning strategy is proposed to address unlabeled and scalable problems on large-scale graphs. Contrastive learning-based methods which can capture both local and global information, compare the mutual information between positive and negative local node representations and global graph representations on the latent space.

3) Experimental results show that GAMMI performs competitively on node classification benchmarks, achieving good results and sometimes even surpassing the performance of supervised learning.

## 2 Related work

In studies on natural language processing, the vector representation of literature is obtained by training largescale pre-trained language models, including Word2Vec [30]based on word context prediction, ELMo [31]based on contextual Word Embedding bidirectional dynamic adjustment, and Transformer[32] based bidirectional language model BERT[33][34]. An important issue is how to build a suitable neural network structure for certain specialized tasks in the research of representation learning based on deep learning method. Convolutional neural networks, recurrent neural networks[35] and attention mechanism are primary methods.

Petar[36] proposed a graph attention network, which uses a masked self-attention layer to solve the shortcomings of previous methods based on graph convolution or its approximation.

Empirical research shows that, through deep neural network learning features, representation learning can have strong data representation capabilities[37][38], and can learn more general prior knowledge independent of a specific task.

At present, there are three main categories of methods in the field of graph embedding based on factorization, random walks, and deep learning. Yu proposed a factorization-based text representation algorithm[39], where they created matrices that each measure a pair of examples' similarities in two different ways. Kawin proposes a random walk model[40], in which the probability of a word being generated is inversely correlated with the angular distance between the word and sentence embeddings. In order to match and rank text for relevant information, a semantic representation method based on CNN is proposed by Zhou[41].

Since most scientific literature data are unlabeled, this makes unsupervised graph-based semantic representation learning crucial. Unsupervised graph learning[42] mainly relies on random walk objectives, which is highly dependent on the choice of parameters. Contrastive methods are at the core of many popular word embedding methods, and Yuning[43] proposed a graph contrast learning framework to study the impact of various combinations of graph enhancement on multiple data sets, which can generate graphical representations with similar or better versatility, portability and robustness.

## 3 The proposed GAMMI method

In this section, this paper proposes an unsupervised semantic representation learning method for scientific literature based on graph attention mechanism and maximum mutual information (GAMMI). First, the graph attention encoder is utilized to learn representation for nodes, and then an unsupervised graph contrastive learning strategy is utilized to solve the problem of being unlabeled and scalable on large-scale graphs. The framework of GAMMI method is shown in Figure 1.

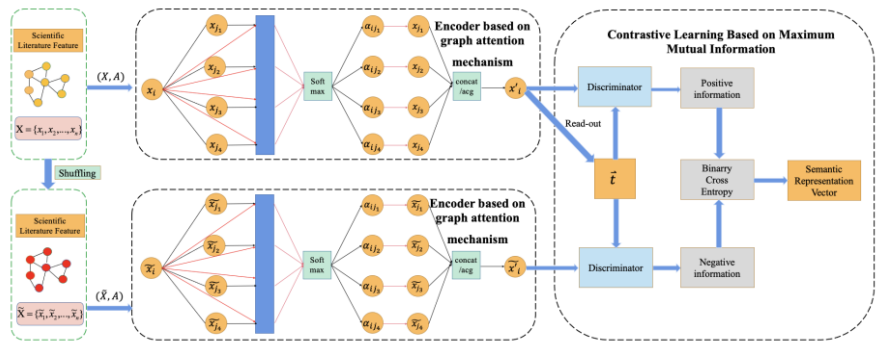

**Figure 1:** The framework of the proposed GAMMI method.

### 3.1 Encoder based on graph attention mechanism

This paper utilized linear transformation to convert the input characteristics into higher-level features. In the graph attention layer, first we use a weight matrix $W \in R^{F' \times F}$ to impact on each literature, and then utilize self-attention to calculate an attention coefficient. The shared self-attention mechanism is expressed as attention coefficients $a$: $R^{F'} \times R^{F'} \to R$

$$e_{ij} = a(W\vec{x_i}, W\vec{x_j}) \quad (1)$$

where $e_{ij}$ is the significance of node $j$'s features to node $i$. We introduce softmax regularizes for all $i$'s neighbors $j$:

$$\alpha_{ij} = softmax_j(e_{ij}) = \frac{exp(e_{ij})}{\sum_{m \in N_i} exp(e_{im})} \quad (2)$$

After obtaining the weight matrix between the connection layers of neural network, we use LeakyReLu function to the output layer of the feedforward neural network.

Combining the above formulas (1) and (2), the complete attention mechanism can be obtained as follows:

$$\alpha_{ij} = \frac{exp(LeakyReLu(\vec{a}^T[W\vec{x_i} \| W\vec{x_j}]))}{\sum_{m \in N_i} exp(LeakyReLu(\vec{a}^T[W\vec{x_i} \| W\vec{x_m}]))} \quad (3)$$

where $\alpha_{ij}$ and $e_{ij}$ are both called attention coefficients. $\alpha_{ij}$ is obtained after softmax normalization based on $e_{ij}$.

The regularized attention coefficients between different nodes are obtained through the above operations. We use it to predict the output features of each literature:

$$\vec{x_i'} = PReLu(\sum_{j \in N_i} \alpha_{ij} W \vec{x_j}) \quad (4)$$

where $W$ is the weight matrix multiplied by the features, $\alpha$ is the attention cross-correlation coefficient calculated earlier, and $j \in N_i$ represents all nodes adjacent to $i$.

To steady the process of self-attention learning, it is beneficial to use multi-head attention. Specifically, Eq.4 is transformed by introducing $K$ independent attention mechanisms, then $K$-average operations are performed on their features, and the final PReLU function is applied. It can be expressed mathematically as:

$$\vec{x_i'} = PReLu(\frac{1}{K}\sum_{k=1}^{K}\sum_{j \in N_i} \alpha_{ij}^k W^k \vec{x_j}) \quad (5)$$

where $K$ represents the number of attention heads, $\alpha^k$ represents the $k$-th attention mechanism, and $W^k$ reflects the input feature's linear transformation weight matrix under the $k$-th attention mechanism.

### 3.2 Contrastive Learning Based on Maximum Mutual Information

Graph attention layer produces node embedding, $\vec{x_i'}$, which summarizes graph patch centered around node $i$, not just the node itself. To get graph-level summary vector $\vec{t}$, this paper uses the readout function, $F: R^{N \times F} \to R^F$ which obtains the feature representation of the entire graph by aggregating node features. The process is represented as:

$$\vec{t} = F(E(X, A)) \quad (6)$$

The readout function $F$ can be a straightforward permutation-invariant function. This paper uses following readout function to get graph-level representation:

$$F(X) = \frac{1}{N}\sum_{i=1}^{N} \vec{x_i'} \quad (7)$$

As a metric to maximize local mutual information, this paper uses a discriminator, $D: R^F \times R^F \to R$. $D(\vec{x_i}, \vec{t})$ is the probability scores given to this patch-summary group.

$$D(\vec{x_i}, \vec{t}) = \sigma(\vec{x_i}^T W \vec{t}) \quad (8)$$

In this paper, we use the random shuffle function to generate graphs with negative samples. This process is keeping the adjacency matrix unchanged, and randomly scrambling the characteristic matrix by row.

In this paper, a discriminator is used to measure noise by contrast, and we use the binary cross-entropy to calculate loss. The discriminator can accurately differentiate between negative and positive samples to enhance the JS divergence which increases the mutual information between local feature representation and global graph representation. The following formula expresses the above process:

$$\mathcal{L} = \frac{1}{N+M}\left(\sum_{i=1}^{N}\mathbb{E}_{(X,A)}\left[\log \mathcal{D}(\vec{x_i}, \vec{t})\right] + \sum_{j=1}^{M}\mathbb{E}_{(\tilde{X},\tilde{A})}\left[\log(1 - \mathcal{D}(\vec{\tilde{x}_j}, \vec{t}))\right]\right) \quad (9)$$

## 4 Experimental results and analysis

### 4.1 Datasets

This paper utilizes Cora and Citeseer datasets[29] to evaluate the proposed GAMMI algorithm. In datasets, the nodes represent literature and the edges represent the connection between two adjacent nodes. Table I shows an overview of the datasets.

**Table I** Summary of the datasets used in experiments.

|  | Cora | Citeseer |
|---|---|---|
| **Node** | 2708 | 3327 |
| **Edges** | 5429 | 4732 |
| **Features/Node** | 1433 | 3703 |
| **Classes** | 7 | 6 |
| **Training Nodes** | 140 | 120 |
| **Validation Nodes** | 500 | 500 |
| **Test Nodes** | 1000 | 1000 |

### 4.2 Experiment 1: Comparative experimental analysis

In the experiment, accuracy, Macro-F1 and recall are adopted to evaluate the performance of GAMMI method for scientific literature classification tasks.

The comparative experimental results are shown in Table II, which indicates that GAMMI is very effective. On all datasets, the performance of GAMMI is the best of all methods. On Cora and Cites, the performance of GAMMI is 1.6% and 1.6% better than that of GCN respectively, which shows that it is effective to differentiate weights to nodes in the same neighborhood. The main reason for this result is that the weighted summation of nearby node features makes the weights of adjacent node features entirely depend on the node features and independent of the graph structure., while GCN is limited to two levels of neighborhood.

**Table II:** Comparative experimental results.

| Model | Core | | | Citeseer | | |
|---|---|---|---|---|---|---|
| | Accuracy | Macro-F1 | Recall | Accuracy | Macro-F1 | Recall |
| DeepWalk | 0.673 | 0.652 | 0.667 | 0.443 | 0.416 | 0.423 |
| MLP | 0.561 | 0.515 | 0.503 | 0.465 | 0.455 | 0.451 |
| GAE | 0.783 | 0.756 | 0.758 | 0.578 | 0.569 | 0.562 |
| VGAE | 0.775 | 0.767 | 0.764 | 0.564 | 0.557 | 0.533 |
| GCN | 0.815 | 0.781 | 0.788 | 0.703 | 0.686 | 0.632 |
| MoNet | 0.811 | 0.779 | 0.787 | 0.691 | 0.672 | 0.611 |
| GraphSAGE | 0.819 | 0.780 | 0.786 | 0.713 | 0.695 | 0.655 |
| GAMMI | **0.831** | **0.801** | **0.799** | **0.719** | **0.708** | **0.687** |

### 4.3 Experiment 2: Ablation experimental analysis

The GAMMI model consists of two components, one is a graph attention encoder, and the other is based on a contrastive learning model based on maximize mutual information.

**Table III** Ablation experimental analysis.

| Model | Core | | | Citeseer | | |
|---|---|---|---|---|---|---|
| | Accuracy | Macro-F1 | Recall | Accuracy | Macro-F1 | Recall |
| GAMMI-attention | 0.825 | 0.789 | 0.786 | 0.671 | 0.653 | 0.621 |
| GAMMI-contrastive | 0.731 | 0.687 | 0.653 | 0.462 | 0.398 | 0.387 |
| GAMMI | **0.831** | **0.801** | **0.799** | **0.719** | **0.708** | **0.687** |

In order to verify the effectiveness of each component, these two components are removed separately, GAMMI-attention only uses the graph attention encoder to learn the semantic representation of scientific literature data, and GAMMI-contrastive maximizes the mutual information of positive and negative samples based on contrastive learning, so as to carry out the semantic representation of scientific literature data. Then we evaluate the performance of different variants. Table III shows the main results. According to the experimental results, it is finally found that GAMMI performs better than GAMMI-attention and GAMMI-contrastive, proving that these components are effective.

## 5 Conclusions

This paper proposes unsupervised semantic representation learning model of scientific literature based on graph attention mechanism and maximum mutual information (GAMMI). Through the introduction of graph attention mechanism, different weights are given to the nodes, and the characteristics of adjacent nodes are weighted and summed. In addition, an unsupervised graph contrastive learning strategy is utilized. By comparing the mutual information between the positive and negative local node representations on the latent space and the global graph representation, the graph neural network can capture both local and global information. Experimental results indicates that GAMMI achieves competitive performance in scientific literature classification tasks, sometimes even better than some supervised architectures.

## Acknowledgements

This work was supported by the National Natural Science Foundation of China (No.62192784, No.62172056, No.61877006).

## References

[1] R. M. Aidi Ahmi, "Bibliometric analysis of global scientific literature on web accessibility," Nternational Journal of Recent Technology and Engineering (IJRTE), vol. 7, no. 6, pp. 250–258, 2019.

[2] Ang Li, Junping Du, Feifei Kou, Zhe Xue, Xin Xu, Mingying Xu, Yang Jiang. "Scientific and Technological Information Oriented Semantics-adversarial and Media-adversarial Cross-media Retrieval," arXiv preprint arXiv:2203.08615, 2022.

[3] M. Ionescu, A. Sterca, and I. Badarinza, "Syntactic indexes for text retrieval," IT in Industry, International Journal of Computer Applications, vol. 5, 2017.

[4] D. Yan, K. Li, S. Gu, and L. Yang, "Network-based bag-of-words model for text classification," IEEE Access, vol. 8, pp. 82641–82652, 2020.

[5] S. Sulova, L. Todoranova, B. Penchev, and R. Nacheva, "Using text mining to classify research papers," in 17th International Multidisciplinary Scientific GeoConference SGEM 2017, 2017.

[6] Mingxing Li, Yinmin Jia, and Junping Du. "LPV control with decoupling performance of 4WS vehicles under velocity-varying motion," IEEE Transactions on Control Systems Technology 2014, 22(5): 1708-1724.

[7] Zeli Guan, Yawen Li, Zhe Xue, Yuxin Liu, Hongrui Gao, Yingxia Shao. "Federated Graph Neural Network for Cross-graph Node Classification," In 2021 IEEE 7th International Conference on Cloud Computing and Intelligent Systems, 418-422, 2021

[8] Wenling Li, Yingmin Jia, and Junping Du. "Distributed extended Kalman filter with nonlinear consensus estimate," Journal of the Franklin Institute, 2017, 354(17): 7983-7995.

[9] Wenling Li, Yingmin Jia, and Junping Du. "Tobit Kalman filter with time-correlated multiplicative measurement noise," IET Control Theory & Applications, 2016, 11(1): 122-128.

[10] O. Shahmirzadi, A. Lugowski, and K. Younge, "Text similarity in vector space models: a comparative study," in 2019 18th IEEE international conference on machine learning and applications (ICMLA). IEEE, 2019, pp. 659–666.

[11] Zeyu Liang, Junping Du, and Chaoyang Li. "Abstractive social media text summarization using selective reinforced Seq2Seq attention model," Neurocomputing, 410 (2020): 432-440.

[12] C. Sun, X. Qiu, Y. Xu, and X. Huang, "How to fine-tune bert for text classification?" in China national conference on Chinese computational linguistics. Springer, 2019, pp. 194–206.

[13] Wenling Li, Jian Sun, Yingmin Jia, Junping Du, and Xiaoyan Fu. "Variance-constrained state estimation for nonlinear complex networks with uncertain coupling strength," Digital Signal Processing, 2017, 67: 107-115.

[14] Liang Xu, Junping Du, Qingping Li. "Image fusion based on nonsubsampled contourlet transform and saliency-motivated pulse coupled neural networks," Mathematical Problems in Engineering, 2013.

[15] Wenling Li, Yingmin Jia, Junping Du. "Distributed consensus extended Kalman filter: a variance-constrained approach," IET Control Theory & Applications, 11(3): 382-389, 2017.

[16] Deyuan Meng, Yingmin Jia, and Junping Du. "Consensus seeking via iterative learning for multi-agent systems with switching topologies and communication time-delays," International Journal of Robust and Nonlinear Control, 2016, 26(17): 3772-3790.

[17] W. Wang, Z. Gan, W. Wang, D. Shen, J. Huang, W. Ping, S. Satheesh, and L. Carin, "Topic compositional neural language model," in International Conference on Artificial Intelligence and Statistics. PMLR, 2018, pp. 356–365.

[18] Peng Lin, Yingmin Jia, Junping Du, Fashan Yu. "Average consensus for networks of continuous-time agents with delayed information and jointly-connected topologies," 2009 American Control Conference, 2009: 3884-3889.

[19] Deyuan Meng, Yingmin Jia, Junping Du, and Fashan Yu, "Tracking Algorithms for Multiagent Systems," In IEEE Transactions on Neural Networks and Learning Systems, 2013, 24(10): 1660-1676.

[20] Yawen Li, Ye Yuan, Yishu Wang, Xiang Lian, Yuliang Ma, Guoren Wang. Distributed Multimodal Path Queries. IEEE Transactions on Knowledge and Data Engineering, 34(7):3196-321, 2022.

[21] Yawen Li, Guangcan Tang, Jiameng Du, Nan Zhou, Yue Zhao, Tian Wu. Multilayer perceptron method to estimate real-world fuel consumption rate of light duty vehicles. IEEE Access, 7, 63395-63402, 2019.

[22] Feifei Kou, Junping Du, Congxian Yang, Yansong Shi, Wanqiu Cui, Meiyu Liang, and Yue Geng. "Hashtag recommendation based on multi-features of microblogs," Journal of Computer Science and Technology, 2018, 33(4): 711-726.

[23] Yawen Li, Isabella Yunfei Zeng, Ziheng Niu, Jiahao Shi, Ziyang Wang and Zeli Guan, "Predicting vehicle fuel consumption based on multi-view deep neural network," Neurocomputing, 502:140-147, 2022.

[24] Yawen Li, Fang Liu, Tiannan Zhang, Fang Xu, Yuchen Gao, Tian Wu.. Artificial intelligence in pediatrics. Chinese Medical Journal, 133(03), 358-360, 2020.

[25] S. Wu, F. Sun, W. Zhang, X. Xie, and B. Cui, "Graph neural networks in recommender systems: a survey," ACM Computing Surveys (CSUR), 2020.

[26] Z. Wu, S. Pan, F. Chen, G. Long, C. Zhang, and S. Y. Philip, "A comprehensive survey on graph neural networks," IEEE transactions on Neural Networks and Learning Systems, vol. 32, no. 1, pp. 4–24, 2020.

[27] Yawen Li, Di Jiang, Rongzhong Lian, Xueyang Wu, Conghui Tan, Yi Xu, Zhiyang Su. "Heterogeneous Latent Topic Discovery for Semantic Text Mining," IEEE Transactions on Knowledge and Data Engineering, 2021.

[28] R. Yin, K. Li, G. Zhang, and J. Lu, "A deeper graph neural network for recommender systems," Knowledge-Based Systems, vol. 185, p. 105020, 2019.

[29] M. Khosla, V. Setty, and A. Anand, "A comparative study for unsupervised network representation learning," IEEE Transactions on Knowledge and Data Engineering, vol. 33, no. 5, pp. 1807–1818, 2019.

[30] S. Thavareesan and S. Mahesan, "Sentiment lexicon expansion using word2vec and fasttext for sentiment prediction in tamil texts," in 2020 Moratuwa Engineering Research Conference (MERCon). IEEE, 2020, pp. 272–276.

[31] D. N. Popa, J. Perez, J. Henderson, and E. Gaussier, "Implicit discourse relation classification with syntax-aware contextualized word representations," in The Thirty-Second International Flairs Conference, 2019.

[32] J. Devlin, M.-W. Chang, K. Lee, and K. Toutanova, "Bert: Pre-training of deep bidirectional transformers for language understanding," arXiv preprint arXiv:1810.04805, 2018.

[33] I. Beltagy, K. Lo, and A. Cohan, "Scibert: A pretrained language model for scientific text," arXiv preprint arXiv:1903.10676, 2019.

[34] Xunpu Yuan, Yawen Li, Zhe Xue, Feifei Kou. "Financial sentiment analysis based on pre-training and textcnn," Chinese Intelligent Systems Conference, 48-56, 2020.

[35] R. Wang, Z. Li, J. Cao, T. Chen, and L. Wang, "Convolutional recurrent neural networks for text classification," in 2019 International Joint Conference on Neural Networks (IJCNN). IEEE, 2019, pp. 1–6.

[36] Petar Veličković, Guillem Cucurull, Arantxa Casanova et al. Graph Attention Networks. In ICLR. 2018.

[37] S. Minaee, N. Kalchbrenner, E. Cambria, N. Nikzad, M. Chenaghlu, and J. Gao, "Deep learning– based text classification: a comprehensive review," ACM Computing Surveys (CSUR), vol. 54, no. 3, pp. 1–40, 2021.

[38] Yingxia Shao, Shiyue Huang, Yawen Li, Xupeng Miao, Bin Cui, Lei Chen. Memory-aware framework for fast and scalable second-order random walk over billion-edge natural graphs.

[39] Z. Wu, S. Pan, F. Chen, G. Long, C. Zhang, and S. Y. Philip, "A comprehensive survey on graph neural networks," IEEE transactions on Neural Networks and Learning Systems, vol. 32, no. 1, pp. 4–24, 2020.

[40] K. Ethayarajh, "Unsupervised random walk sentence embeddings: A strong but simple baseline," in Proceedings of the Third Workshop on Representation Learning for NLP, 2018, pp. 91–100.

[41] Nan Z, Junping Du, Xu Yao, et al. "Microblog Topic Content Search Method Based on Convolutional Neural Networks". Journal of Frontiers of Computer Science & Technology, 2019, 13(5): 753.

[42] R. Sato, "A survey on the expressive power of graph neural networks," arXiv preprint arXiv:2003.04078, 2020.

[43] Yuning You, Tianlong Chen, Yongduo Sui et al. Graph Contrastive Learning with Augmentations. In NeurIPS. 2020.